\documentclass[10pt]{scrartcl}
\pdfoutput=1

\usepackage[T1]{fontenc}
\usepackage[utf8]{inputenc}

\usepackage{amsmath,amsfonts}
\usepackage[mathscr]{euscript}
\usepackage{booktabs}
\usepackage{cite}
\usepackage{graphicx, subfigure, ulem, hhline}
\usepackage[table]{xcolor}
\usepackage{siunitx}
\usepackage{soul}
\usepackage{wasysym}              % adds \permil
\usepackage{hyphenat}             % advanced hyphenation control

\usepackage{etoolbox,xstring,xspace}

\let\theparentequation\theequation
\patchcmd{\theparentequation}{equation}{parentequation}{}{}

\renewenvironment{subequations}[1][]{
  \refstepcounter{equation}%
  \setcounter{parentequation}{\value{equation}}%    parentequation = equation
  \setcounter{equation}{0}%                         (sub)equation  = 0
  \def\theequation{\theparentequation\alph{equation}}% 
  \let\parentlabel\label%                           Evade sanitation performed by amsmath
  \ifx\\#1\\\relax\else\label{#1}\fi%               #1 given: \label{#1}, otherwise: nothing
  \ignorespaces
}{%
  \setcounter{equation}{\value{parentequation}}%    equation = subequation
  \ignorespacesafterend
}

\newcommand*{\nextParentEquation}[1][]{
  \refstepcounter{parentequation}%                  parentequation++
  \setcounter{equation}{0}%                         equation = 0
  \ifx\\#1\\\relax\else\parentlabel{#1}\fi%         #1 given: \label{#1}, otherwise: nothing
}

\makeatletter
\renewcommand\@makefntext[1]{\leftskip=.7em\hskip-.68em\@makefnmark#1}
\makeatother

\usepackage{setspace}

\usepackage{jheppub}
\usepackage[all]{hypcap}
\usepackage{footnotebackref}

\usepackage[format=plain, margin=0.cm, font=small, labelfont=bf]{caption}
\DeclareCaptionSubType*[arabic]{figure}
\DeclareCaptionSubType*[arabic]{table}
\let\theparentequation\theequation
\patchcmd{\theparentequation}{equation}{parentequation}{}{}

\apptocmd{\thebibliography}{\small}{}{}
\let\OLDthebibliography\thebibliography
\renewcommand\thebibliography[1]{
  \OLDthebibliography{#1}
  \setlength{\parskip}{0pt}
  \setlength{\itemsep}{1.5pt plus 0.3ex}
}

\def\unity{\mathbf{1}}

\newcommand{\IE}{\textit{i.\,e.}}
\newcommand{\EG}{\textit{e.\,g.}}
\newcommand{\I}{{i\mkern1mu}}
\newcommand{\E}{\mathrm{e}}
\newcommand{\Real}[1]{\Re\hspace{-1pt}\mathfrak{e}{\left[#1\right]}}

\newcommand{\MZ}{\mathbf{M}^{\mathbf{Z}}}

\newcommand{\sgn}[1]{\text{sgn}{\left[#1\right]}}

\makeatletter
\DeclareRobustCommand\em{%
  \@nomath\em \ifdim \fontdimen\@ne\font >\z@\scshape
  \else \slshape \fi}
\makeatother

\renewcommand{\emph}[1]{{\em #1}}

\hypersetup{pdfauthor={Sebastian Pa{\ss}ehr and Georg Weiglein},
            pdftitle={Two-loop top and bottom Yukawa corrections to the Higgs-boson masses in the complex MSSM}}

\begin{document}

% title page and abstract

\thispagestyle{empty}

\def\thefootnote{\fnsymbol{footnote}}

\begin{flushright}
DESY--17--065%\\
%arXiv:yymm.nnnn [hepph]
\end{flushright}

\vspace{2cm}

\begin{center}

{\large\textsc{\textbf{ Two-loop top and bottom Yukawa corrections to the}}}

\vspace{0.4cm}

{\large\textsc{\textbf{ Higgs-boson masses in the complex MSSM}}}

\vspace{1cm}

Sebastian Pa{\ss}ehr\footnote{email: sebastian.passehr@desy.de}
and
Georg Weiglein\footnote{email: georg.weiglein@desy.de}

\vspace*{.7cm}

\textsl{
Deutsches Elektronensynchrotron DESY\\
Notkestraße 85, D--22607 Hamburg, Germany
}

\end{center}

\vspace*{2cm}

\begin{abstract}{}
  Results for the two-loop corrections to the Higgs-boson masses of
  the MSSM with complex parameters of
  $\mathcal{O}{\left(\alpha_t^2+\alpha_t\alpha_b+\alpha_b^2\right)}$
  from the Yukawa sector in the gauge-less limit are presented.  The
  corresponding self-energies and their renormalization have been
  obtained in the Feynman-diagrammatic approach. The impact of the new
  contributions on the Higgs spectrum is investigated. Furthermore, a
  comparison with an existing result in the limit of the MSSM with
  real parameters is carried out. The new results will be included in
  the public code \texttt{FeynHiggs}.
\end{abstract}
%\pacs{}

\def\thefootnote{\arabic{footnote}}
\setcounter{page}{1}
\setcounter{footnote}{0}

% 1. Introduction
\section{Introduction}

After the discovery of a Higgs
boson~\cite{Aad:2012tfa,Chatrchyan:2012ufa} with a mass
around~$125$\,GeV, intense studies were performed to reveal its
nature. Although within the present experimental uncertainties the
measured properties of this new boson are consistent with the
expectations for the Higgs boson of the Standard Model
(SM)~\cite{Oda:2017,Khachatryan:2016vau}, it could be part of an
extended model like the theoretically well motivated minimal
supersymmetric Standard Model~(MSSM). In the MSSM the observed
particle could in principle be interpreted as one of the three neutral
physical Higgs bosons. At the tree-level, the physical states are
given by the neutral $CP$-even $h,H$ and $CP$-odd $A$ bosons, together
with the charged $H^{\pm}$ bosons, and can be parametrized in terms of
the $A$-boson mass $m_A$ and the ratio of the two vacuum expectation
values, $\tan\beta = \left.v_2\middle/v_1\right.$. An admixture of
these $CP$ eigenstates is introduced to the Higgs sector via loop
contributions involving complex parameters from other supersymmetric
(SUSY)
sectors~\cite{Pilaftsis:1998pe,Demir:1999hj,Pilaftsis:1999qt,Heinemeyer:2001qd}.
 
Loop corrections to the masses of the Higgs bosons are sizable and
therefore phenomenologically very important. Accordingly, numerous
calculations for higher-order corrections to the mass spectrum within
the MSSM for the case where $CP$ conservation has been
assumed~\cite{Haber:1990aw,Ellis:1990nz,Okada:1990vk,Okada:1990gg,Ellis:1991zd,Sasaki:1991qu,Chankowski:1991md,Brignole:1992uf,Hempfling:1993qq,Casas:1994us,Dabelstein:1994hb,Carena:1995bx,Carena:1995wu,Pierce:1996zz,Haber:1996fp,Heinemeyer:1998kz,Zhang:1998bm,Heinemeyer:1998jw,Heinemeyer:1999be,Heinemeyer:1998np,Degrassi:2002fi,Espinosa:1999zm,Carena:2000dp,Espinosa:2000df,Espinosa:2001mm,Degrassi:2001yf,Martin:2001vx,Brignole:2001jy,Dedes:2002dy,Martin:2002iu,Dedes:2003km,Martin:2003qz,Martin:2003it,Allanach:2004rh,Heinemeyer:2004gx,Martin:2005eg,Martin:2005qm,Harlander:2008ju,Kant:2010tf,Harlander:2017kuc,Brignole:2002bz,Heinemeyer:2004xw,Borowka:2014wla,Degrassi:2014pfa,Borowka:2015ura,Draper:2013oza,Vega:2015fna,Lee:2015uza,Bahl:2016brp,Hahn:2013ria,Athron:2016fuq,Bahl:2017aev,Martin:2002wn,Martin:2004kr,Martin:2007pg}
as well as for the general case of the MSSM with complex
parameters~(cMSSM)~\cite{Pilaftsis:1998pe,Demir:1999hj,Pilaftsis:1999qt,Choi:2000wz,Ibrahim:2000qj,Ibrahim:2002zk,Carena:2000yi,Heinemeyer:2001qd,Martin:2002wn,Martin:2004kr,Frank:2006yh,Martin:2007pg,Heinemeyer:2007aq,Hollik:2014wea,Hollik:2014bua,Goodsell:2016udb}
have already been performed.  The largest one-loop contributions
originate from the Yukawa sector due to the size of the top-quark
Yukawa coupling~$h_t$,
where~\mbox{$\alpha_t=\left.h_t^2\middle/(4\pi)\right.$}. For large
values of $\tan\beta$ contributions of the order
$\alpha_b=\left.h_b^2\middle/(4\pi)\right.$, with the bottom Yukawa
coupling $h_b$, can become sizable. At the two-loop level both types
of contributions receive further potentially large corrections. The
dominant contribution is given by the leading
$\mathcal{O}{\left(\alpha_{t}\alpha_{s}\right)}$
terms~\cite{Heinemeyer:1998jw,Heinemeyer:1998np,Heinemeyer:1999be,Heinemeyer:2007aq}
which are known in the MSSM with complex parameters. Additional
corrections involving the strong coupling~$\alpha_s$ are known in the
special case of the $CP$-conserving
MSSM~\cite{Brignole:2002bz,Heinemeyer:2004xw,Borowka:2014wla,Degrassi:2014pfa,Borowka:2015ura}. Another
important class of two-loop corrections are Yukawa-coupling enhanced
contributions of the
order~$\mathcal{O}{\left(\alpha_t^2+\alpha_t\alpha_b+\alpha_b^2\right)}$
which are known in the $CP$-conserving MSSM as
well~\cite{Brignole:2001jy,Dedes:2003km} (the corrections
of~$\mathcal{O}{\left(\alpha_t\alpha_b+\alpha_b^2\right)}$ with
on-shell parameters are only available in the
approximation~$\tan\beta\to\infty$ and~$m_b\to 0$). A computation of
the leading corrections of~$\mathcal{O}{\left(\alpha_t^2\right)}$ has
been published for the general
MSSM~\cite{Hollik:2014wea,Hollik:2014bua}. In this article also the
other pieces of the two-loop Yukawa terms are obtained for the general
case of the MSSM with complex parameters.

The phases of complex parameters in the MSSM are constrained by limits
on electric dipole
moments~(EDMs)~\cite{Baron:2013eja,Pospelov:2005pr,Afach:2015sja,Baker:2006ts,Baker:2007df,Serebrov:2015idv},
the impact of meson mixings and decays (see
Ref.~\cite{Goodsell:2015yca} and references therein), and
Higgs-coupling measurements~\cite{Khachatryan:2016vau}.

Following the usual convention, we choose to fix the phase of the mass
of the electroweakinos,~$\phi_{M_2}$, to zero; then the phase of~$\mu$
from the superpotential,~$\phi_\mu$, needs to be close to zero
or~$\pi$ in order to be compatible with the experimental
constraints. The other relevant parameters are the phase of the gluino
mass parameter,~$\phi_{M_3}$, and the trilinear soft-breaking
parameters of the stops,~$\phi_{A_t}$, and sbottoms,~$\phi_{A_b}$.
These phases,~$\phi_{M_3}$, $\phi_{A_t}$ and~$\phi_{A_b}$, are less
constrained; especially the bounds on the phases of the trilinear
soft-breaking parameters are weaker for the third generation than for
the second and first generation.

\medskip

The calculation presented here extends the Yukawa-type contributions
of $\mathcal{O}{\left(\alpha_t^2\right)}$ from
Refs.~\cite{Hollik:2014wea,Hollik:2014bua}, and profits from
previously developed tools~\cite{Hahn:2015gaa}. For this reason the
theoretical framework is just briefly outlined and only new aspects
are explained in detail in Section~\ref{sec:HiggsSect}. The numerical
analysis in Section~\ref{sec:numeric} is focused on the impact of the
new contributions on the Higgs masses, showing substantial shifts of
$2$\,GeV and more in certain regions of parameter space. In the limit
of vanishing phases of the parameters, our results agree well with
previous results in the MSSM for the case where $CP$ conservation has
been assumed~\cite{Dedes:2003km}; differences are shown for the
comparison with an interpolation for non-zero phases which so far has
been used in
\texttt{FeynHiggs}~\cite{Heinemeyer:1998np,Degrassi:2002fi,Frank:2006yh,Heinemeyer:1998yj,Hahn:2010te}. The
new results will become part of the public code \texttt{FeynHiggs}.

\section{Higgs masses at higher orders in the complex MSSM\label{sec:HiggsSect}}

In this section, we briefly outline the theoretical framework for
Higgs-mass predictions at higher orders in the MSSM. We introduce our
conventions, explain some details of our chosen renormalization
scheme, and comment on the gauge-less limit and the bottom mass
resummation.

\subsection{Notation and conventions at the tree level}

The two scalar $SU(2)$ Higgs doublets are expressed in terms of their
components in the following way,
\begin{alignat}{2}
  \label{eq:Higgsfields}
  \mathcal{H}_{1} &= \begin{pmatrix} v_{1} + \frac{1}{\sqrt{2}}(\phi_{1} - \I \chi_{1})\\ -\phi^{-}_{1}\end{pmatrix},&\quad
  \mathcal{H}_{2} &= \begin{pmatrix} \phi^{+}_{2}\\ v_{2} + \frac{1}{\sqrt{2}}(\phi_{2} + \I \chi_{2})\end{pmatrix} .
\end{alignat}
After rotation to mass eigenstates, the Higgs potential reads
\begin{align}
\label{eq:HiggsPotential}
  \begin{split}
    V_{H} &= -T_h \, h- T_H \, H - T_A \, A - T_G\,  G\\
         &\quad + \frac{1}{2}\begin{pmatrix} h, & H, & A, & G \end{pmatrix}
            \mathbf{M}_{hHAG}
            \begin{pmatrix} h \\ H\\ A\\ G\end{pmatrix}
            + \begin{pmatrix} H^{-}, & G^{-}\end{pmatrix} 
                \mathbf{M}_{H^\pm G^\pm} 
                \begin{pmatrix} H^{+}\\ G^{+}\end{pmatrix} + \dots\ ,
  \end{split}
\end{align}
with the tadpole coefficients $T_{h,H,A,G}$, and the mass matrices
\begin{align}
\label{eq:mmatrices}
\mathbf{M}_{hHAG} &= \begin{pmatrix}
        m^2_{h} & m^2_{hH} & m^2_{hA} & m^2_{hG} \\
        m^2_{hH} & m^2_{H} & m^2_{HA} & m^2_{HG} \\
        m^2_{hA} & m^2_{HA} & m^2_{A} & m^2_{AG} \\
        m^2_{hG} & m^2_{HG} & m^2_{AG} & m^2_{G} \end{pmatrix} , \qquad
\mathbf{M}_{H^\pm G^\pm} \,= \begin{pmatrix}
        m^2_{H^\pm}  &  m^2_{H^-G^+} \\
        m^2_{G^-H^+} &  m^2_{G^\pm} \end{pmatrix}.
\end{align}
The matrices $\mathbf{M}_{hHAG}$ and $\mathbf{M}_{H^\pm G^\pm}$ are
diagonal at the tree level after minimizing the potential. Explicit
expressions for the entries are given in Ref.~\cite{Frank:2006yh}.

\subsection{Gauge-less limit}

The gauge-less limit in our calculation is defined by neglecting all
couplings proportional to~$g_1$ or~$g_2$. As a consequence of this
approximation the gauge-boson masses~$M_W$ and~$M_Z$ are equal to zero
in the new two-loop contributions.

Accordingly, the Higgs-boson masses entering the two-loop calculation
take on the values
\begin{align}
  m_h &= m_G = m_{G^\pm} = 0\,,& m_H &= m_A = m_{H^\pm}\,.
\end{align}
In this limit, the tree-level mixing
angles~$\alpha\in\left[-\pi/2,0\right)$
  and~$\beta\in\left[0,\pi/2\right)$ fulfill the relation
\begin{align}
  \alpha &= \beta - \frac{\pi}{2}\,.
\end{align}

\subsection{Higgs masses at the two-loop order}

The Higgs mass matrix elements at the two-loop order receive
contributions from self-energies, leading in general to mixing of all
neutral states. In this article the full one-loop corrections are
used, while the $\mathcal{O}{\left(\alpha_t\alpha_s\right)}$ and the
new $\mathcal{O}{\left(\alpha_t^2+\alpha_t\alpha_b+\alpha_b^2\right)}$
terms are evaluated in the gauge-less limit and at zero external
momentum. Therefore, the loop-corrected propagator~$\Delta_{h H A G}$
is given by
\begin{align}
  \label{eq:masscorr}
    \Delta_{h H A G}(p^2) &= \I\left[p^2\unity - \mathbf{M}_{ h H A G}^{(0)} + \mathbf{\hat{\Sigma}}_{h H A G}^{(1)}(p^2) + \mathbf{\hat{\Sigma}}_{h H A G}^{(2)}(0)\right]^{-1}\, .
\end{align}
Therein, $ \mathbf{\hat{\Sigma}}_{h H A G} ^{(k)}$ denotes the matrix
of the renormalized diagonal and non-diagonal self-energies for the
$h, H, A, G$ fields at loop order $k$, and $\mathbf{M}_{ h H A
  G}^{(0)}$ denotes the diagonal tree-level mass matrix.

Mixing of the Goldstone boson (and of the longitudinal $Z$ boson) with
the other Higgs bosons yields negligible effects to the propagators of
the physical Higgs
bosons~\cite{Baro:2008bg,Williams:2011bu,Hollik:2002mv}. Therefore, in
the following we will only consider the~$\left(3\times 3\right)$
submatrix of $\Delta_{h H A G}$ involving the physical Higgs
bosons. Though, Goldstone--Higgs mixing is taken into account in
subloop renormalization terms of the type
$\left(\text{one-loop}\right)^2$~\cite{Hollik:2014bua}.

The neutral Higgs masses are derived from the real parts of the
complex poles of the $hHA$ propagator matrix, obtained as the zeros
of the determinant of the renormalized two-point function,
\begin{align}
  \label{eq:higgspoles}
   \operatorname{det}\hat{\Gamma}_{hHA}{\left(p^2\right)} &= 0\,, &
   \hat{\Gamma}_{hHA}{\left(p^2\right)} &= \I \left[p^2 {\unity} - \mathbf{M}_{ h H A}^{(0)} + \mathbf{\hat{\Sigma}}_{h H A}^{(1)}(p^2) + \mathbf{\hat{\Sigma}}_{h H A}^{(2)}(0)\right].
\end{align}

\subsection{Counterterms}

The renormalized two-loop self-energies can be written as
\begin{align}
\label{eq:renselfenergies}
\mathbf{\hat{\Sigma}}_{h H A} ^{(2)} (p^2)  
   & =
\mathbf{\Sigma}_{h H A} ^{(2)} (p^2) - \delta^{(2)} \MZ_{hHA} \, ,
\end{align}
with $\mathbf{\Sigma}_{h H A} ^{(2)}$ denoting the unrenormalized
self-energies at the two-loop order, and $\delta^{(2)} \MZ_{hHA}$
comprising all two-loop counterterms resulting from parameter and
field renormalization. The notation follows~\cite{Hollik:2014bua},
where the required expressions for~$\delta^{(2)} \MZ_{hHA}$ can be
found.

The Feynman-diagrammatic calculation of the self-energies has been
performed with the help
of~\texttt{FeynArts}~\cite{Kublbeck:1990xc,Hahn:2000kx} for the
generation of the Feynman diagrams, and
\texttt{TwoCalc}~\cite{Weiglein:1993hd} for the two-loop tensor
reduction and trace evaluation. The one-loop renormalization constants
have been obtained with the help of
\texttt{FormCalc}~\cite{Hahn:1998yk}.

\subsubsection{Genuine two-loop renormalization}

The two-loop counterterms for the Higgs self-energies given in
Ref.~\cite{Hollik:2014bua} also apply to the corrections described in
the present article. However, there is an interesting difference for
the cancelation of the divergence in the self-energy
$\Sigma^{(2)}_{hH}(0)$. The corresponding counterterm reads
\begin{align}\label{eq:mhH}
  \delta^{(2)}m_{hH}^{\mathbf{Z}} &= \delta^{(2)}m_{hH}^2 + \tfrac{1}{2}m_{H^{\pm}}^2\delta^{(2)}Z_{hH} + \dots\,,
\end{align}
where terms with products of two one-loop counterterms have been
omitted. In the gauge-less limit $\delta^{(2)}m_{hH}^2$ is the only
counterterm which contains $\delta^{(2)}t_{\beta}$,
\begin{align}
  \delta^{(2)}m_{hH}^2 &= m_{H^{\pm}}^2\,c_{\beta}^2\,\delta^{(2)}t_{\beta} + \dots\,.
\end{align}
Here we define $t_\beta\equiv\tan\beta$, $s_\beta\equiv\sin\beta$ and
$c_\beta\equiv\cos\beta$. The two-loop field renormalization constant
for the same matrix element is given by
\begin{align}
  \delta^{(2)}Z_{hH} &= -c_{\beta}s_{\beta}\left[\delta^{(2)}Z_{\mathcal{H}_2} - \tfrac{1}{4}\left(\delta^{(1)}Z_{\mathcal{H}_2}\right)^2 - \delta^{(2)}Z_{\mathcal{H}_1} + \tfrac{1}{4}\left(\delta^{(1)}Z_{\mathcal{H}_1}\right)^2\right]\,.
\end{align}
The two-loop counterterm for $t_{\beta}$ in the $\overline{\text{DR}}$
scheme and in the gauge-less limit can be expressed as
\begin{align}\label{eq:tb}
  \delta^{(2)}t_{\beta} &= \frac{t_{\beta}}{2}\left[\left(\delta^{(2)}Z_{\mathcal{H}_2} - \delta^{(2)}Z_{\mathcal{H}_1}\right) - \tfrac{1}{4}\left(\delta^{(1)}Z_{\mathcal{H}_2} - \delta^{(1)}Z_{\mathcal{H}_1}\right)^2 -\left(\delta^{(1)}Z_{\mathcal{H}_2} - \delta^{(1)}Z_{\mathcal{H}_1}\right) \delta^{(1)}Z_{\mathcal{H}_1}\right]\,.
\end{align}
Combining Eqs.~\eqref{eq:mhH}--\eqref{eq:tb} yields
\begin{align}\label{eq:mhHZ}
  \delta^{(2)}m_{hH}^{\mathbf{Z}} &= \frac{c_{\beta}\,s_{\beta}\,m_{H^{\pm}}^2}{4}\left(\delta^{(1)}Z_{\mathcal{H}_2} - \delta^{(1)}Z_{\mathcal{H}_1}\right)\delta^{(1)}Z_{\mathcal{H}_1} + \dots\,.
\end{align}
The $\overline{\text{DR}}$ field-renormalization
constant~$\delta^{(1)}Z_{\mathcal{H}_1}$ is a pure UV-divergent term,
calculated as the derivative with respect to the external
momentum~$p^2$ of the $\phi_1$ Higgs self-energy. The only
contribution in the gauge-less limit is a bottom loop, \IE~in the case
of the previously calculated $\mathcal{O}{\left(\alpha_t^2\right)}$
corrections~\cite{Hollik:2014bua}, $\delta^{(1)}Z_{\mathcal{H}_1}$ was
equal to zero due to the approximation $m_b=0$. The terms originating
from two-loop field-renormalization and two-loop renormalization of
$t_{\beta}$ canceled each other exactly.

Now, for the more general case of a non-zero bottom mass, also
$\delta^{(1)}Z_{\mathcal{H}_1}$ is non-zero and the cancelation is not
complete anymore. The genuine two-loop parts of the
field-renormalization constants, $\delta^{(2)}Z_{\mathcal{H}_1}$ and
$\delta^{(2)}Z_{\mathcal{H}_2}$, drop out in the gauge-less limit at
zero external momentum in Eq.~\eqref{eq:mhHZ} because of a cancelation
of the contributions in~$\delta^{(2)}t_\beta$
and~$\delta^{(2)}Z_{hH}$. In principle,
$\delta^{(2)}Z_{\mathcal{H}_1}$ and $\delta^{(2)}Z_{\mathcal{H}_2}$
could still appear as field-renormalization constants for the other
Higgs-mass counterterms. However, also there they drop out exactly
(see Eq.~(2.23) in~\cite{Hollik:2014bua}):
\begin{itemize}
\item for $\delta^{(2)}m_h^{\mathbf{Z}}$ since $m_h^2 = 0$ in the gauge-less limit,
\item for $\delta^{(2)}m_H^{\mathbf{Z}}$ since $m_H^2 = m_{H^\pm}^2$ and $\alpha = \beta - \tfrac{\pi}{2}$ in the gauge-less limit,
\item for $\delta^{(2)}m_A^{\mathbf{Z}}$ since $m_A^2 = m_{H^\pm}^2$ in the gauge-less limit,
\item for $\delta^{(2)}m_{hA}^{\mathbf{Z}}$ and $\delta^{(2)}m_{HA}^{\mathbf{Z}}$ since the Higgs sector is $CP$ conserving at the tree level.
\end{itemize}

\subsubsection{Resummation}

Radiative corrections to the relation between the bottom-quark mass
and the Yukawa coupling of the bottom quark~$h_b$ are proportional to
$t_{\beta}$. In order to resum the leading $t_\beta$-enhanced
contributions, an effective bottom Yukawa coupling is used as
described in
Refs.~\cite{Banks:1987iu,Hall:1993gn,Hempfling:1993kv,Carena:1994bv,Carena:1999py,Eberl:1999he,Williams:2011bu},
leading to a UV finite and complex correction factor $\Delta
m_b$. Using a $\overline{\text{DR}}$ renormalization for~$m_b$ in
the~MSSM, the largest contributions of this type are captured through
an effective bottom-quark mass which is given by
\begin{align}\label{eq:deltamb}
  m_b^{\overline{\text{DR}},\text{MSSM}}(m_t^{\text{os}}) \simeq m_{b,\text{eff}} &= \frac{m_b^{\overline{\text{DR}},\text{SM}}(m_t^{\text{os}})}{\lvert 1 + \Delta m_b \rvert}.
\end{align}
The symbol $m_b^{\overline{\text{DR}},\text{SM}}(m_t^{\text{os}})$
denotes the bottom mass in the $\overline{\text{DR}}$ renormalization
scheme, taking into account SM-type QCD corrections, evaluated at the
on-shell top mass.

We use the correction factor $\Delta m_b$ at the one-loop order which
is implemented in \texttt{FeynHiggs}. For illustrating the effects seen
in our numerical analysis below, we give here the explicit form of the
leading contributions:
\begin{subequations}\label{eq:db}
\begin{align}
  \Delta m_b &= \frac{2\,\alpha_s}{3\,\pi}\,\mu^*\,M_3^*\,t_\beta\,\mathcal{I}{\left(m_{\tilde{b}_1}^2,m_{\tilde{b}_2}^2,m_{\tilde{g}}^2\right)}
  +\frac{\alpha_t}{4\,\pi}\,\mu^*\,A_t^*\,t_\beta\,\mathcal{I}{\left(m_{\tilde{t}_1}^2,m_{\tilde{t}_2}^2,\lvert\mu\rvert^2\right)}\,,\\
  \mathcal{I}{\left(a,b,c\right)} &=
    -\frac{b\,a\,\log{\tfrac{b}{a}} +
           c\,b\,\log{\tfrac{c}{b}} +
           a\,c\,\log{\tfrac{a}{c}}}
    {\left(b - a\right)\left(c - b\right)\left(a - c\right)}.
\end{align}
\end{subequations}
Further subleading contributions involve terms~$\propto\alpha_b$
and~$\propto\alpha$. With this definition a part of the considered
two-loop corrections
of~$\mathcal{O}{\left(\alpha_t\alpha_b+\alpha_b^2\right)}$ to the
Higgs-boson masses is absorbed into an effective bottom-quark mass. In
order to avoid a double counting of contributions from the
bottom--sbottom sector to the Higgs-boson self-energies, the
bottom-mass is renormalized in the $\overline{\text{DR}}$ scheme as
specified in Eq.~\eqref{eq:deltamb}.

\subsubsection{Subloop renormalization}

One-loop counterterms for subloop renormalization enter the
self-energies $\mathbf{\Sigma}_{h H A} ^{(2)}$ in
Eq.~\eqref{eq:renselfenergies}. In contrast to the previously
calculated $\mathcal{O}{\left(\alpha_t^2\right)}$ corrections, the
approximation of massless bottom quarks is dropped in the present
calculation. Accordingly, new counterterms for the bottom--sbottom
sector are induced, which are specified in the following.

The squark mass matrices in the
$\big(\tilde{q}_{\text{L}},\tilde{q}_{\text{R}}\big)$ bases, $q=t,b$,
in the gauge-less limit are given by
\begin{align}
  \label{eq:squarks}
    \mathbf{M}_{\tilde{q}} &= 
    \begin{pmatrix}
     m_{\tilde{q}_{\text{L}}}^{2} + m_{q}^{2} & 
     m_{q}\left(A_{q}^{*} - \mu\,\kappa_q \right)\\[0.1cm]
     m_{q}\left(A_{q} - \mu^{*}\,\kappa_q \right) & 
     m_{\tilde{q}_{\text{R}}}^{2} + m_{q}^{2}
   \end{pmatrix}, &
     \kappa_{t} &= \frac{1}{t_{\beta}},& \kappa_{b} &= t_{\beta} .
\end{align}
$SU(2)$-invariance requires \mbox{$m_{\tilde{t}_{\text{L}}}^{2} =
  m_{\tilde{b}_{\text{L}}}^{2} \equiv m_{\tilde{Q}_{3}}^{2}$}.  The squark mass
eigenvalues can be obtained by performing unitary transformations,
\begin{align}
  \label{eq:squarkdiag}
  \mathbf{U}_{\tilde{q}}\mathbf{M}_{\tilde{q}}\mathbf{U}_{\tilde{q}}^{\dagger}  &= 
  \mathrm{diag}{\left(m_{\tilde{q}_{1}}^{2},\, m_{\tilde{q}_{2}}^{2}\right)}.
\end{align}
The independent parameters entering the two-loop calculation via the
quark--squark sector are the quark masses $m_{q}$, the soft
SUSY-breaking parameters $m_{\tilde{Q}_{3}}$ and
$m_{\tilde{q}_{\text{R}}}$, $\tilde{q} = \tilde{t},\tilde{b}$, the
complex trilinear couplings~\mbox{$A_{q} = \lvert
  A_{q}\rvert\E^{i\,\phi_{A_{q}}}$, $q=t,b$}, the complex $\mu$
parameter from the superpotential, and the ratio of the vacuum
expectation values $t_{\beta}$. All of them have to be renormalized at
the one-loop level,
\begin{align}\label{eq:squarkren}
  m_q &\rightarrow m_q + \delta^{(1)}m_q\ , &
  \mathbf{M}_{\tilde{q}}  &\rightarrow \mathbf{M}_{\tilde{q}} + \delta^{(1)}\mathbf{M}_{\tilde{q}}\ .
\end{align}
Here~$\delta^{(1)}\mathbf{M}_{\tilde{q}}$ denotes the matrix of
counterterms after applying the renormalization transformation to the
parameters in Eq.~\eqref{eq:squarks}. The renormalization of the
top--stop sector, as well as of $\mu$ and $t_{\beta}$ is carried out
as specified in Ref.~\cite{Hollik:2014bua}.

For the renormalization of the bottom--sbottom sector, we refer to
Refs.~\cite{Dedes:2003km,Heinemeyer:2004xw,Heinemeyer:2010mm} where
renormalization of $m_b$ and $A_b$ in the $\overline{\text{DR}}$
scheme has been proposed to avoid numerical instabilities. Also for
the applied resummation of~$\Delta m_b$ the $\overline{\text{DR}}$
scheme for~$m_b$ is convenient, as explained above. The
renormalization scale is chosen to be the on-shell top mass.
\begin{itemize}
\item The bottom-quark self-energy in Lorentz decomposition is given
  by
\begin{align}
\label{eq:Lorentz} 
\Sigma_b (p) & =\, \not{\! p}\, \omega_-\, \Sigma_b^{\mathrm{L}} (p^2) +
\not{\! p}\, \omega_+\, \Sigma_b^{\mathrm{R}}(p^2) + m_b \,\Sigma_b^{\mathrm{S}}(p^2) + m_b \gamma_5\, \Sigma_b^{\mathrm{PS}}(p^2)\,,
\end{align}
with the left-vector part $\Sigma_b^{\mathrm{L}}$, right-vector part
$\Sigma_b^{\mathrm{R}}$, scalar part $\Sigma_b^{\mathrm{S}}$, and pseudo-scalar
part $\Sigma_b^{\mathrm{PS}}$. The bottom-quark mass renormalization is
fixed at the on-shell top-mass scale via
\begin{align}
  \delta^{(1)}m_{b} &= m_{b} \,
  \Real{\frac{1}{2}\left(\Sigma_{b}^{\text{L}}{\left(m_{b}^{2}\right)}
      + \Sigma_{b}^{\text{R}}{\left(m_{b}^{2}\right)}\right) +
    \Sigma_{b}^{\text{S}}{\left(m_{b}^{2}\right)}}_{\overline{\text{DR}}}.
\end{align}
\item With Eqs.~\eqref{eq:squarkdiag}--\eqref{eq:squarkren} we define
\begin{align}
  \label{eq:stopcountermmatrix}     
  \mathbf{U}_{\tilde{q}}\, \delta\mathbf{M}_{\tilde{q}}\,
  \mathbf{U}_{\tilde{q}}^{\dagger} & =
  \begin{pmatrix} 
    \delta^{(1)}m_{\tilde{q}_{1}}^{2} & \delta^{(1)}m_{\tilde{q}_{1}\tilde{q}_{2}}^{2}  \\
    \delta^{(1)}m_{\tilde{q}_{1}\tilde{q}_{2}}^{2\,*} &
    \delta^{(1)}m_{\tilde{q}_{2}}^{2}
  \end{pmatrix} .
\end{align}
The renormalization of the soft-breaking parameter $A_b$ follows from
Eqs.~\eqref{eq:squarks} and~\eqref{eq:stopcountermmatrix} with $q=b$,
yielding
\begin{align}
  \label{eq:Abrenormalization}
  \begin{split}
    \delta^{(1)}A_{b} &=
    \Bigg[\mathbf{U}_{\tilde{b}\,11}\mathbf{U}_{\tilde{b}\,12}^{*}\frac{\delta^{(1)}M_{\tilde{b}_{1}}^{2} - \delta^{(1)}M_{\tilde{b}_{2}}^{2}}{m_b}
    + \mathbf{U}_{\tilde{b}\,21}\mathbf{U}_{\tilde{b}\,12}^{*}\frac{\delta^{(1)}M_{\tilde{b}_{1}\tilde{b}_{2}}^{2}}{m_b}\\
    &\qquad + \mathbf{U}_{\tilde{b}\,11}\mathbf{U}_{\tilde{b}\,22}^{*}\frac{\delta^{(1)}M_{\tilde{b}_{1}\tilde{b}_{2}}^{2\,*}}{m_b} - \left(A_{b} - \mu^{*}\,t_\beta\right)\frac{\delta^{(1)}m_{b}}{m_b} + \mu^*\,\delta t_\beta + t_\beta\,\delta\mu^*\Bigg]_{\overline{\text{DR}}}\ .
  \end{split}
\end{align}
The divergent parts of the
counterterms~$\delta^{(1)}M_{\tilde{b}_{i}}^{2}$, $i=1,2$, which are
needed for the~$\overline{\text{DR}}$ renormalization in
Eq.~\eqref{eq:Abrenormalization}, can be computed from the
corresponding sbottom
self-energies~$\Sigma^{(1)}_{\tilde{b}_i\tilde{b}_i}$, $i=1,2$, and
the divergent part for the
counterterm~$\delta^{(1)}M_{\tilde{b}_{1}\tilde{b}_{2}}^{2}$ can be
obtained from the sbottom
mixing~$\Sigma^{(1)}_{\tilde{b}_1\tilde{b}_2}$, where the self-energy
is defined with an incoming~$\tilde{b}_2$ and an
outgoing~$\tilde{b}_1$. The renormalization conditions to fix the
auxiliary counterterms~$\delta^{(1)}M_{\tilde{b}_{i}}^{2}$
and~$\delta^{(1)}M_{\tilde{b}_{1}\tilde{b}_{2}}^{2}$ may be chosen
analogously to the stop-sector counterterms in
Ref.~\cite{Hollik:2014bua}. Again, the renormalization scale in
Eq.~\eqref{eq:Abrenormalization} is the on-shell top mass.
\item Invariance under $SU(2)$ yields the following relation between
  the stop and sbottom sector,
\begin{align}\label{eq:SU2_1L}
  \begin{split}
    \delta^{(1)}m_{\tilde{Q}_3}^2 &\equiv \sum\limits_{i\;=\;1}^2\lvert\mathbf{U}_{\tilde{t}\,i1}\rvert^{2}\,\delta^{(1)}m_{\tilde{t}_{i}}^{2} + 2\Real{\mathbf{U}_{\tilde{t}\,21}\mathbf{U}_{\tilde{t}\,11}^{*}\,\delta^{(1)}m_{\tilde{t}_{1}\tilde{t}_{2}}^{2}} - 2\,m_t\,\delta^{(1)}m_t\\
    &= \sum\limits_{i\;=\;1}^2\lvert\mathbf{U}_{\tilde{b}\,i1}\rvert^{2}\,\delta^{(1)}m_{\tilde{b}_{i}}^{2} + 2\Real{\mathbf{U}_{\tilde{b}\,21}\mathbf{U}_{\tilde{b}\,11}^{*}\,\delta^{(1)}m_{\tilde{b}_{1}\tilde{b}_{2}}^{2}} - 2\,m_b\,\delta^{(1)}m_b\ .
  \end{split}
\end{align}
We trade~$\delta^{(1)}m_{\tilde{Q}_3}^2$,
$\delta^{(1)}m_{\tilde{t}_{\text{R}}}^2$ and~$\delta^{(1)}A_t$
for~$\delta^{(1)}m_{\tilde{t}_i}^2$, $i=1,2$,
and~$\delta^{(1)}m_{\tilde{t}_1\tilde{t}_2}^2$, and we apply on-shell
conditions to both stop particles and the stop mixing angle (see
Ref.~\cite{Hollik:2014bua}). Then we choose to
make~$\delta^{(1)}m_{\tilde{b}_{1}}^{2}$ a dependent quantity by the
relation in Eq.~\eqref{eq:SU2_1L}. The other diagonal sbottom-mass
counterterm~$\delta^{(1)}m_{\tilde{b}_{2}}^{2}$ is employed instead of
$\delta^{(1)}m_{\tilde{b}_{\text{R}}}^{2}$ and is fixed
on-shell via
\begin{align}
  \label{eq:sbottomonshell}
  \delta^{(1)}m_{\tilde{b}_{2}}^{2} &= \Real{\, \Sigma_{\tilde{b}_{22}}^{(1)}{\left(m_{\tilde{b}_{2}}^{2}\right)}}.
\end{align}
The quantity $\delta^{(1)}m_{\tilde{b}_{1}\tilde{b}_{2}}^{2}$ is the
off-diagonal entry of Eq.~\eqref{eq:stopcountermmatrix} for $q=b$. It
is already fixed by the renormalization condition of
Eq.~\eqref{eq:Abrenormalization} for the independent
counterterm~$\delta^{(1)}A_b$.
\end{itemize}

\section{Numerical results for the Higgs spectrum}
\label{sec:numeric}

In the following numerical analysis the new contributions of
$\mathcal{O}{\left(\alpha_t^2+\alpha_t\alpha_b+\alpha_b^2\right)}$ are
added to the known Higgs-mass corrections in the general case of the
MSSM with complex parameters which are implemented in
\texttt{FeynHiggs}~(version~\texttt{2.12.0}).\footnote{The previously
  implemented contributions of $\mathcal{O}{\left(\alpha_t^2\right)}$
  are replaced by the new result.} While the improvement by
resummation of leading logarithms as described in
Refs.~\cite{Hahn:2013ria,Bahl:2016brp,Bahl:2017aev} can be applied
also to the case of complex parameters (via an interpolation routine),
we have not included contributions of this kind in our numerical
results presented below.\footnote{For the scenarios in
  Fig.~\ref{fig:stop}, for which in contrast to the other results
  shown below \texttt{FeynHiggs} version~\texttt{2.13.0} has been
  used, the incorporation of higher-order leading and next-to-leading
  logarithmic contributions (\texttt{FeynHiggs} flag
  \texttt{loglevel=1}) would shift the displayed results for~$M_h$ by:
  $0.6$\,GeV~(blue), $1.7$\,GeV~(red) and~$2.7$\,GeV~(green)
  for~$X_t=0$; $-0.1$\,GeV~(blue), $1.0$\,GeV~(red)
  and~$2.0$\,GeV~(green) for $\lvert
  X_t\rvert/m_{\tilde{t}_{\text{R}}}=2$.}  The large impact of the
$\mathcal{O}{\left(\alpha_t^2\right)}$ terms has been investigated in
Refs.~\cite{Hollik:2014wea,Hollik:2014bua} and is not presented here
again. Instead the focus is set on the new corrections induced by the
finite bottom mass. If not stated otherwise, we choose the following
default setting for the parameters entering through the new
contributions:
\begin{subequations}\label{eq:param}
\begin{alignat}{5}
  t_\beta &= 50, &\ m_{H^\pm} &= 1.5\,\text{TeV}, & m_{\tilde{Q}_3} &= 2.1\,\text{TeV}, &\, m_{\tilde{t}_\text{R}} &= m_{\tilde{b}_\text{R}} = 2\,\text{TeV}, &\, m_t &= 173.2\,\text{GeV},\label{eq:real}\\
  A_t &= \left|1.3\,m_{\tilde{t}_\text{R}} + \frac{\mu^*}{t_\beta}\right|\E^{\I\,\phi_{A_t}},\span\ \span & A_b &= 2.5\,m_{\tilde{b}_\text{R}}\,\E^{\I\,\phi_{A_b}}, &\, M_3 &= 2.5\,\text{TeV}\,\E^{\I\,\phi_{M_3}}, &\, \mu &= \sgn{\mu}\,1\,\text{TeV}.\hspace{-1em}\label{eq:complex}
\end{alignat}
\end{subequations}
The quantities in Eq.~\eqref{eq:real} are real parameters. The charged
Higgs mass $m_{H^\pm}$ is chosen as an input parameter, and its value
is set to ensure the compatibility of scenarios with high~$t_\beta$
with the current experimental constraints from searches for heavy MSSM
Higgs bosons~\cite{Khachatryan:2014wca,ATLAS:2016fpj}.  The parameters
in Eq.~\eqref{eq:complex} are in general complex.  Their respective
phases $\phi_{A_t}$, $\phi_{A_b}$ and $\phi_{M_3}$ are scanned in
section~\ref{sec:phases}.  Thereby the gluino mass parameter~$M_3$
does not occur directly in the new Higgs self-energy contributions,
but it appears in the leading term of the bottom-mass resummation. The
parameter~$\mu$ is also complex in general, but its phase is
constrained to be very close to zero or~$\pi$ by EDM limits (see
above). We remark that the phases $\phi_{M_3}$, $\phi_{A_t}$ and
$\phi_{A_b}$ are also constrained by EDM limits, but scenarios with
large phases are possible (see \EG{} Ref.~\cite{Arbey:2014msa}). We
show results for the Higgs mass when varying two phases at the same
time.

The absolute value of~$A_t$ has been fixed to yield a lightest
Higgs-boson mass close to~$125$\,GeV which can then be identified with
the Higgs signal discovered at ATLAS and CMS. Together with
$m_{\tilde{Q}_3}$ and $m_{\tilde{t}_\text{R}}$ it determines the mass
shift which is induced by the stop contributions. We choose different
values for $m_{\tilde{Q}_3}$ and $m_{\tilde{t}_\text{R}}$,
$m_{\tilde{b}_\text{R}}$ to avoid numerical instabilities due to
degeneracies. Different setups for $m_{\tilde{t}_\text{R}}$ and $X_t =
A_t - \mu^*/t_\beta$ are possible to yield a lightest Higgs mass of
$125$\,GeV as can be seen in Fig.~\ref{fig:stop}. Therein the gray bar
indicates the mass range $125.1\pm 0.21(\text{stat})\pm
0.11(\text{syst})$\,GeV as measured by ATLAS and
CMS~\cite{Aad:2015zhl}.

The absolute value of~$A_b$ is close to the upper limit
\begin{subequations}\label{eq:ablimit}
\begin{align}
  \lvert A_b\rvert^2 &< 3 \left(m_{H_d}^2 + \lvert\mu\rvert^2 + m_{\tilde{Q}_3}^2 + m_{\tilde{b}_\text{R}}^2\right),\\
  m_{H_d}^2 + \lvert\mu\rvert^2 &= \left(m_{H^\pm}^2 - m_W^2\right) {\sin^2}\beta - \frac{1}{2} m_Z^2 \cos{2\beta}\ ,
\end{align}
\end{subequations}
from the approximate bound from the requirement of vacuum stability to
avoid charge- and color-breaking
minima~\cite{Frere:1983ag,Gunion:1987qv} (see
Refs.~\cite{Casas:1995pd,Hisano:2006mj,Hisano:2007cz,Hisano:2008hn,Camargo-Molina:2013qva,Camargo-Molina:2014pwa,Hollik:2016dcm}
for more detailed discussions of this issue).

In the following analyses we call $\Delta M_h$ the shift of the
lightest Higgs-boson mass by the new Yukawa terms
of~$\mathcal{O}{\left(\alpha_t\alpha_b+\alpha_b^2\right)}$,
\IE~excluding the previously analyzed contributions
of~$\mathcal{O}{\left(\alpha_t^2\right)}$. In section~\ref{sec:real},
the impact of different parameters on the lightest Higgs boson mass in
the $CP$-conserving case is investigated.

We have also investigated the mass shifts of the heavier neutral Higgs
bosons. In general, the shifts are of the same absolute size as for
the lightest Higgs but with opposite sign. However, since the
tree-level input value $m_{H^\pm}$ needs to be large for high values
of~$t_\beta$ (where the
$\mathcal{O}{\left(\alpha_t\alpha_b+\alpha_b^2\right)}$ contributions
are relevant) to be in agreement with experimental constraints, the
relative mass shift for the heavy Higgs bosons is only $\approx
1\permil$. Moreover, both heavy Higgs bosons receive nearly identical
corrections; in the investigated scenarios the largest difference was
$\approx 0.1$\,GeV. For this reason we do not present numerical
results for the mass shifts of the heavier Higgs bosons here. It
should however be noted that even small mass shifts can have an
important impact on the resonance-type behavior that typically occurs
between the two heavy neutral Higgs states in $CP$-violating
scenarios, see Refs.~\cite{Fuchs:2016swt,Fuchs:2017wkq}.

\subsection{Scenarios with real parameters\label{sec:real}}

\begin{figure}[tp!]
  \centering
  \includegraphics[width=.8\linewidth]{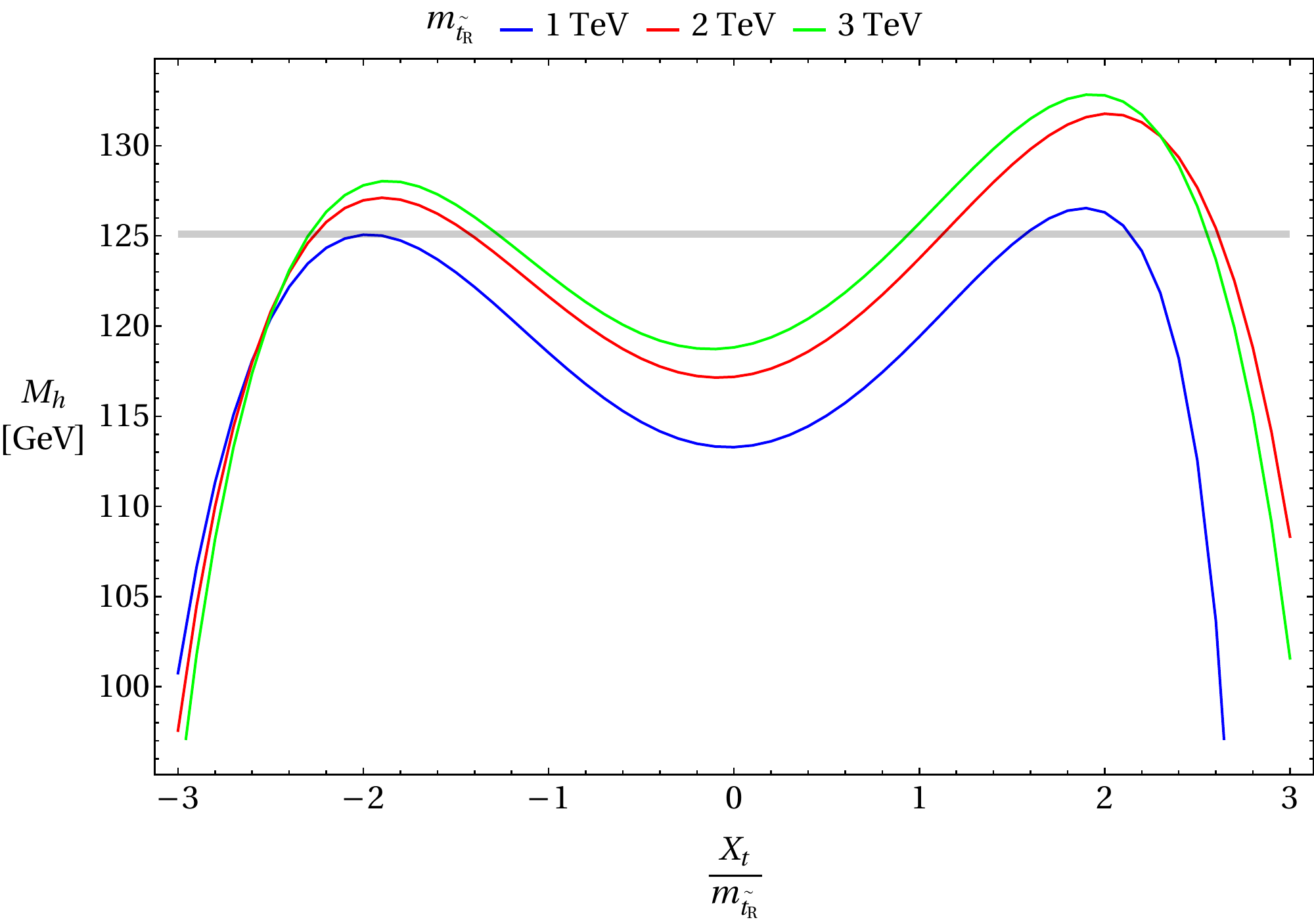}
  \caption{\label{fig:stop}Dependence of the lightest Higgs-mass $M_h$
    on the stop parameters $m_{\tilde{t}_{\text{R}}}$ and
    $X_t=A_t-\mu^*/t_\beta$ as predicted by \texttt{FeynHiggs-2.13.0}
    in the version for real parameters without contributions of
    resummed logarithms. We set
    $m_{\tilde{Q}_3}=m_{\tilde{t}_{\text{R}}}+100$\,GeV. The other
    parameters except $m_{\tilde{Q}_3}$, $m_{\tilde{t}_{\text{R}}}$
    and $A_t$ have been fixed to the values given in
    Eqs.~\eqref{eq:param} with vanishing phases. Two-loop corrections
    of
    $\mathcal{O}{\left(\alpha_t\alpha_s+\alpha_b\alpha_s+\alpha_t^2+\alpha_t\alpha_b+\alpha_b^2\right)}$
    are comprised as implemented in the version of \texttt{FeynHiggs}
    for real parameters.}
\end{figure}

\begin{figure}[bp!]
  \centering
  \includegraphics[width=.8\linewidth]{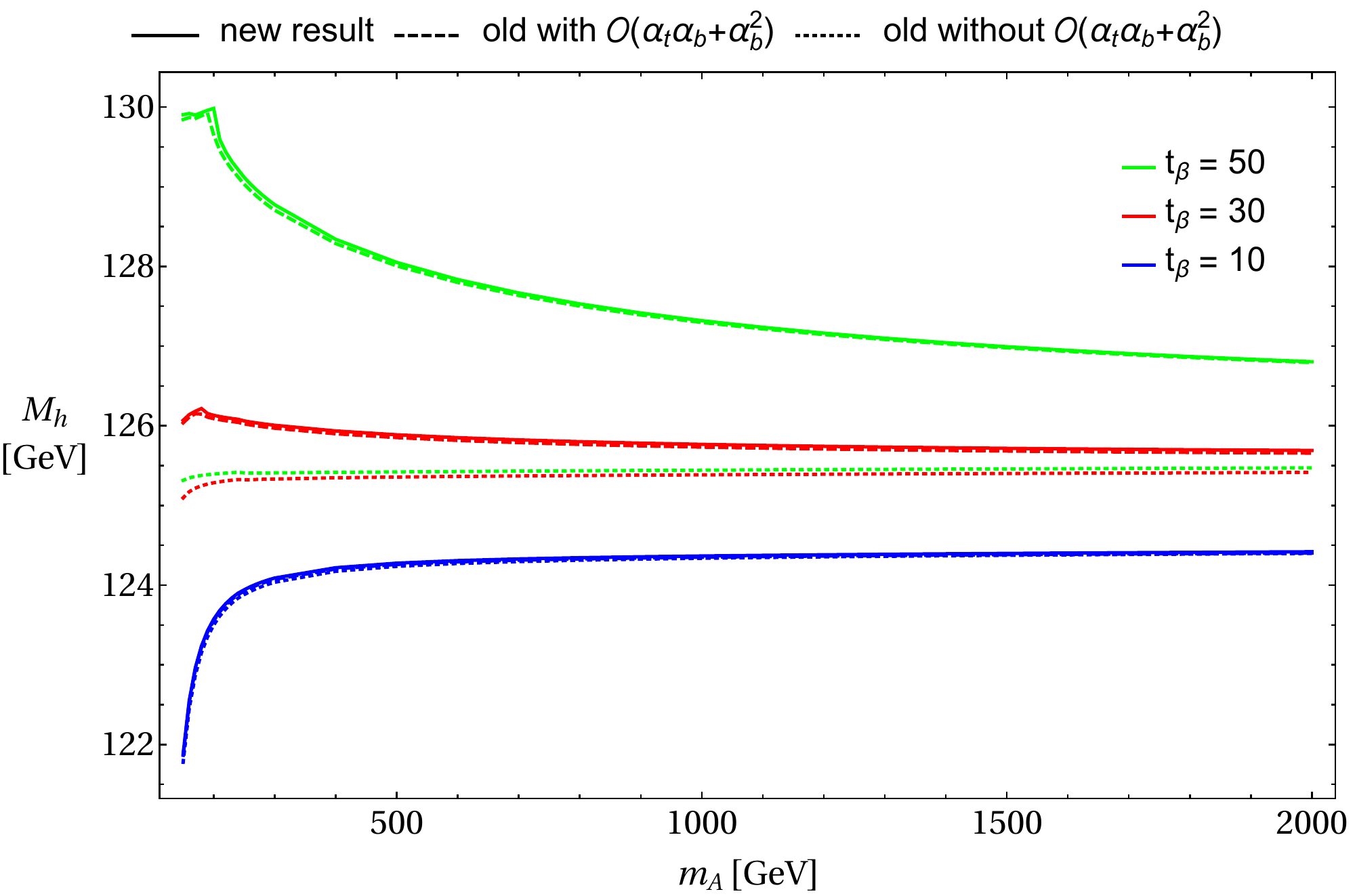}
  \caption{\label{fig:FHcomp}Comparison of the lightest Higgs-boson
    mass $M_h$ as predicted with our new two-loop corrections (solid),
    the version of \texttt{FeynHiggs} for the MSSM with real
    parameters, \IE~including
    $\mathcal{O}{\left(\alpha_t\alpha_b+\alpha_b^2\right)}$
    corrections (dashed), and the version of \texttt{FeynHiggs} for
    the MSSM with complex parameters, \IE~without
    $\mathcal{O}{\left(\alpha_t\alpha_b+\alpha_b^2\right)}$
    corrections (dotted) for the parameters specified in
    Eqs.~\eqref{eq:param} with vanishing phases and~$\sgn{\mu}=-1$.}
\end{figure}

We start to analyze our results by performing a comparison with the
previously implemented two-loop corrections in \texttt{FeynHiggs}. The
two-loop corrections
of~$\mathcal{O}{\left(\alpha_t\alpha_b+\alpha_b^2\right)}$ were up to
now only known for the MSSM with real parameters and~$m_A$ being an
input. We compare our new result with the predictions obtained so far
with~\texttt{FeynHiggs} from both the versions for real
parameters\footnote{The contributions
  of~$\mathcal{O}{\left(\alpha_b\alpha_s\right)}$ beyond~$\Delta m_b$
  are only available in this version and therefore subtracted to allow
  for a clean comparison.} and for complex
parameters (for the latter employing the same renormalization scheme
with~$m_A$ as input and in the limit of real parameters, but without
terms of~$\mathcal{O}{\left(\alpha_t\alpha_b+\alpha_b^2\right)}$).
Both versions contain shifts due to~$\Delta m_b$ effects (see
Eq.~\eqref{eq:db}), including contributions
of~$\mathcal{O}{\left(\alpha_t\alpha_b\right)}$.

In Fig.~\ref{fig:FHcomp} the predictions of~\texttt{FeynHiggs}
(dashed: MSSM with real parameters, dotted: MSSM with complex
parameters) are compared to our new result (solid) as a function of
$m_A$. The different colors correspond to different values of
$t_\beta$. The large deviations between the dashed and dotted curves
for large values of $t_\beta$ are induced by the
$\mathcal{O}{\left(\alpha_t\alpha_b+\alpha_b^2\right)}$ terms, which
are not incorporated in the dotted curve. After adding our new
contributions to the result for complex parameters the agreement with
the~\texttt{FeynHiggs} result for the case of real parameters is very
good, \IE~the dashed and solid lines almost coincide with each other.
Since the version of
the~$\mathcal{O}{\left(\alpha_t\alpha_b+\alpha_b^2\right)}$
corrections which is implemented in~\texttt{FeynHiggs} so far employs
further approximations of~$t_\beta\to\infty$ and~$m_b\to 0$ (see
Ref.~\cite{Dedes:2003km}), while our new result is not simplified
further, the agreement is not expected to be perfect. The largest
difference of~$\approx0.3$\,GeV is found in the threshold region
at~$m_A=m_{\tilde{t}_2}-m_{\tilde{t}_1}\approx200$\,GeV which enters
via the renormalization in the stop sector.

\begin{figure}[b]
  \begin{minipage}[t]{.48\linewidth}
    \includegraphics[width=\linewidth]{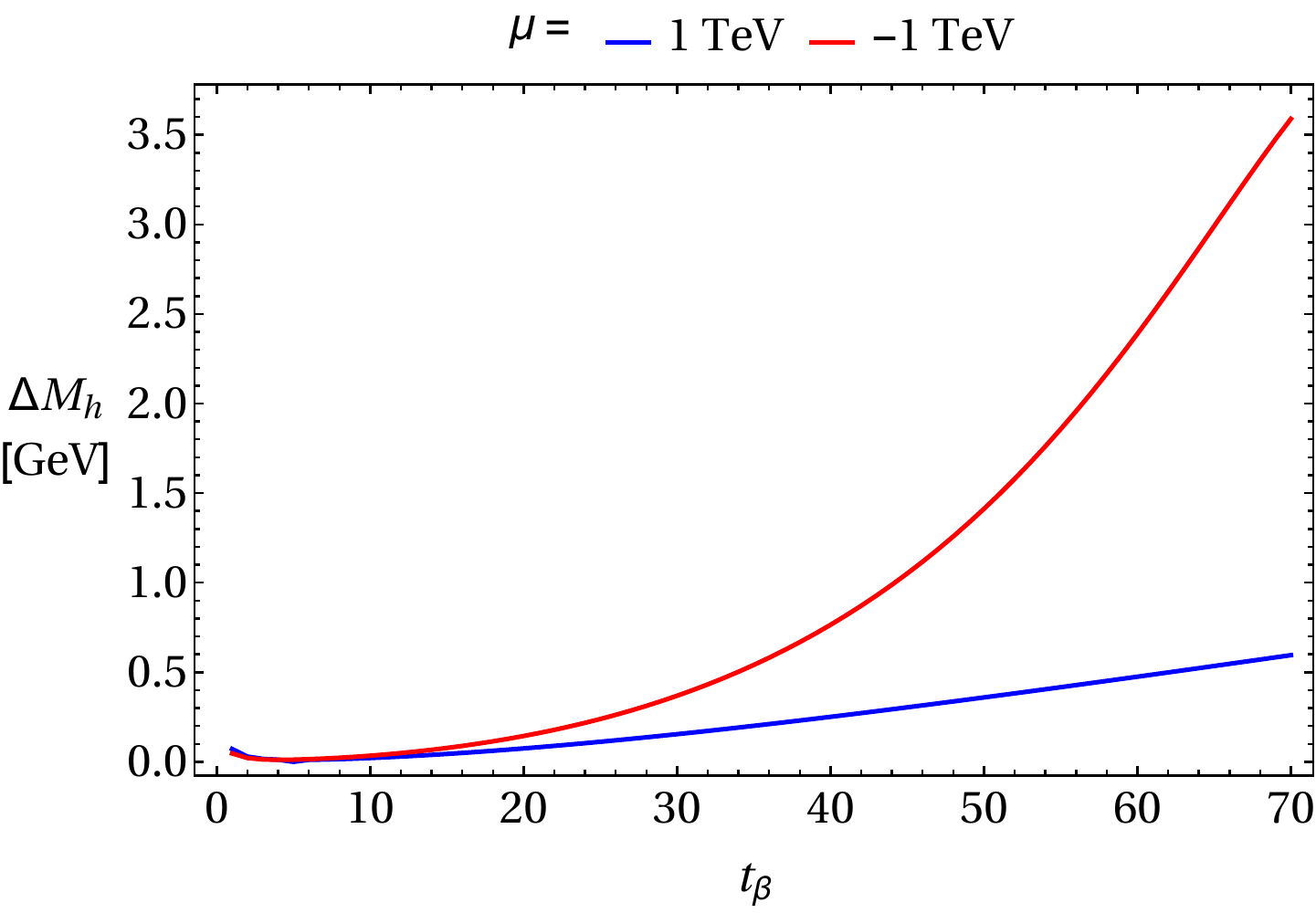}
    \caption{\label{fig:tb}Dependence of the lightest Higgs-mass shift
      $\Delta M_h$ on $t_\beta$. The parameter $\mu$ is either positive
      (blue) or negative (red).}
  \end{minipage}
  \hfill
  \begin{minipage}[t]{.48\linewidth}
    \includegraphics[width=\linewidth]{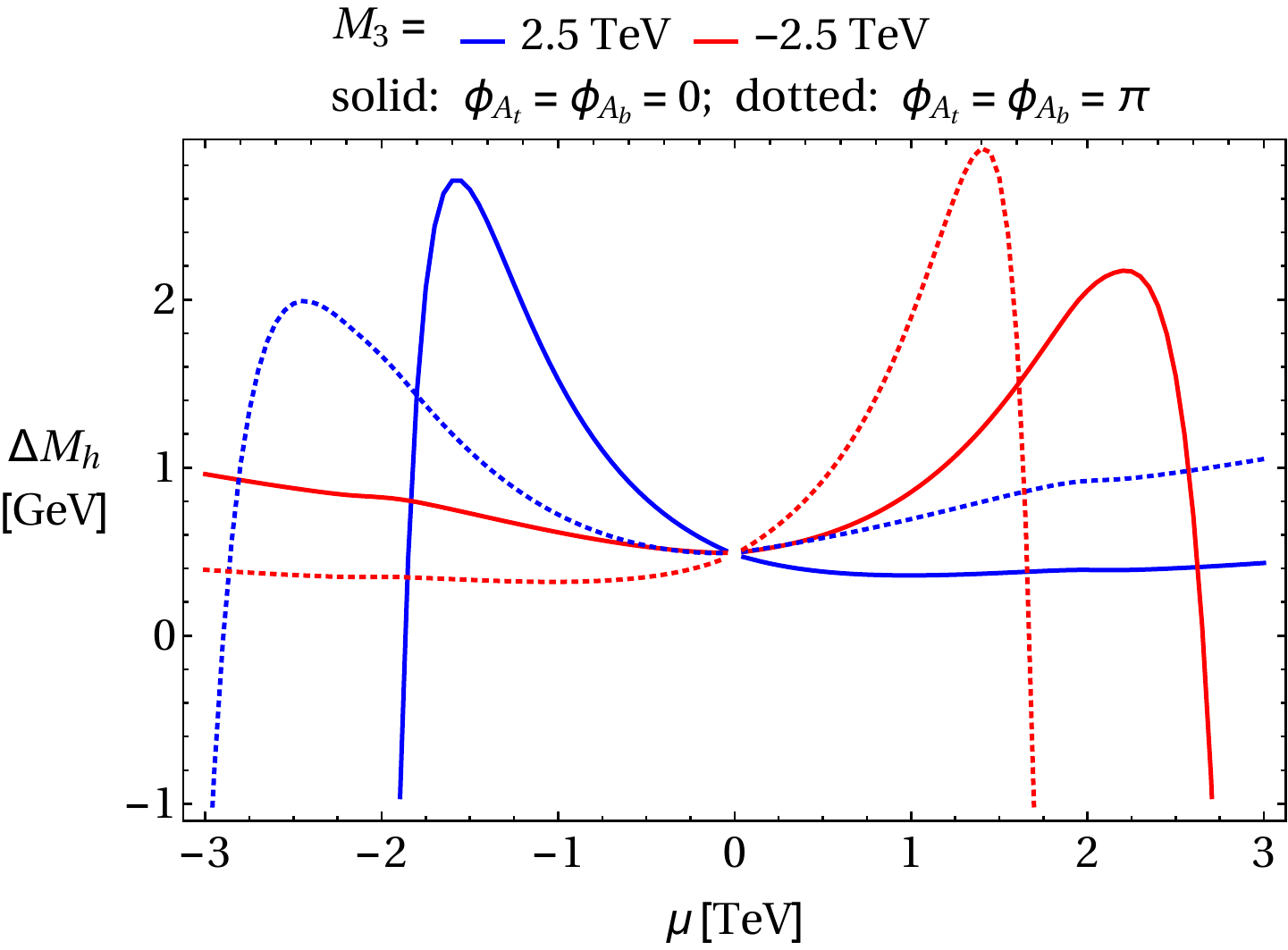}
    \caption{\label{fig:mu}Dependence of the lightest Higgs-mass shift
      $\Delta M_h$ on $\mu$. The parameter $M_3$ is either positive
      (blue) or negative (red). The region around $\mu=0$ is left out
      because of numerical instabilities.}
  \end{minipage}
\end{figure}

\begin{figure}
  \begin{minipage}{.48\linewidth}
  \includegraphics[width=\linewidth]{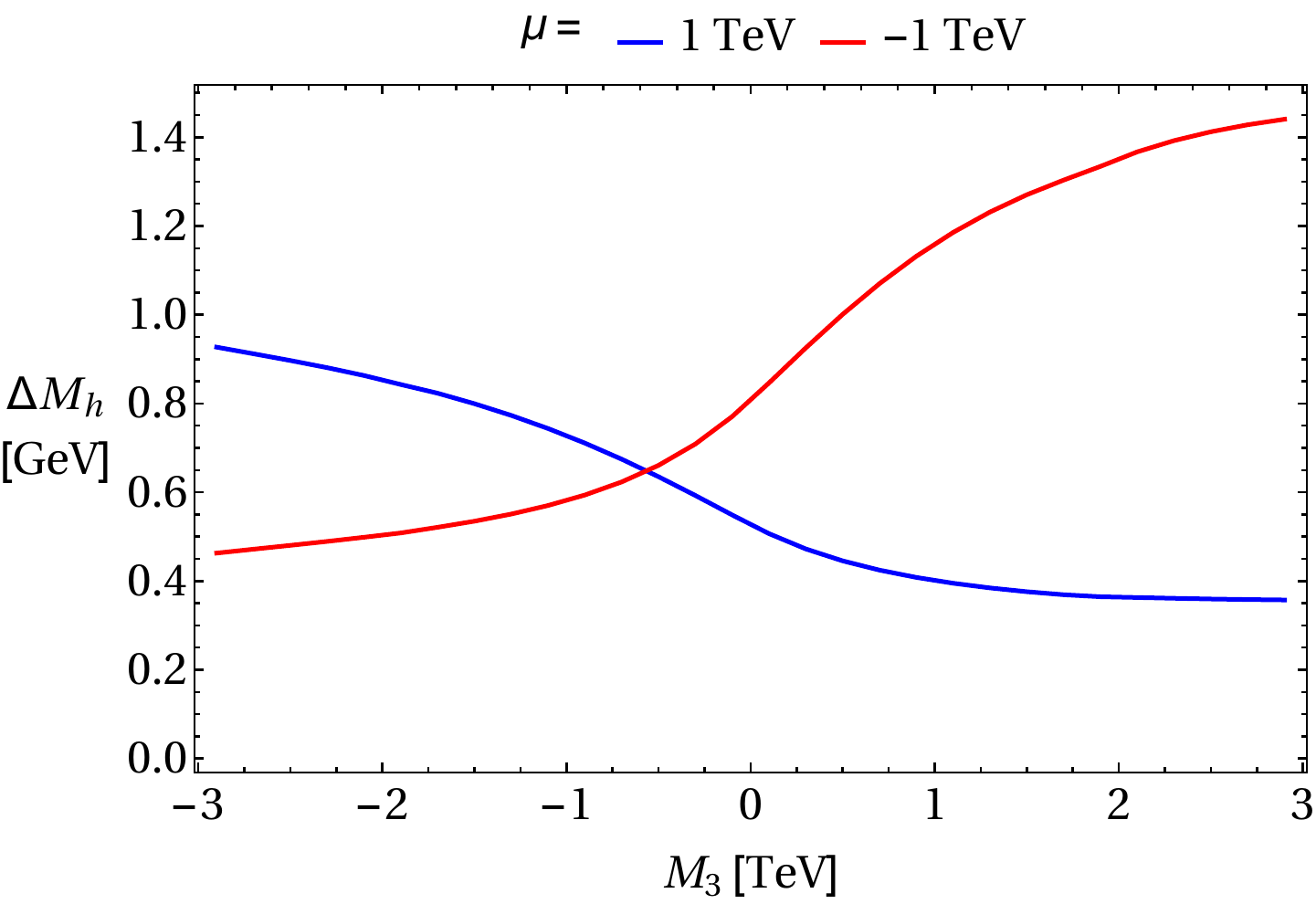}
  \vspace{.4ex}
  \caption{\label{fig:M3}Dependence of the lightest Higgs-mass shift
    $\Delta M_h$ on $M_3$. The parameter $\mu$ is either positive
    (blue) or negative (red).}
  \end{minipage}
  \hfill
  \begin{minipage}{.48\linewidth}
  \includegraphics[width=\linewidth]{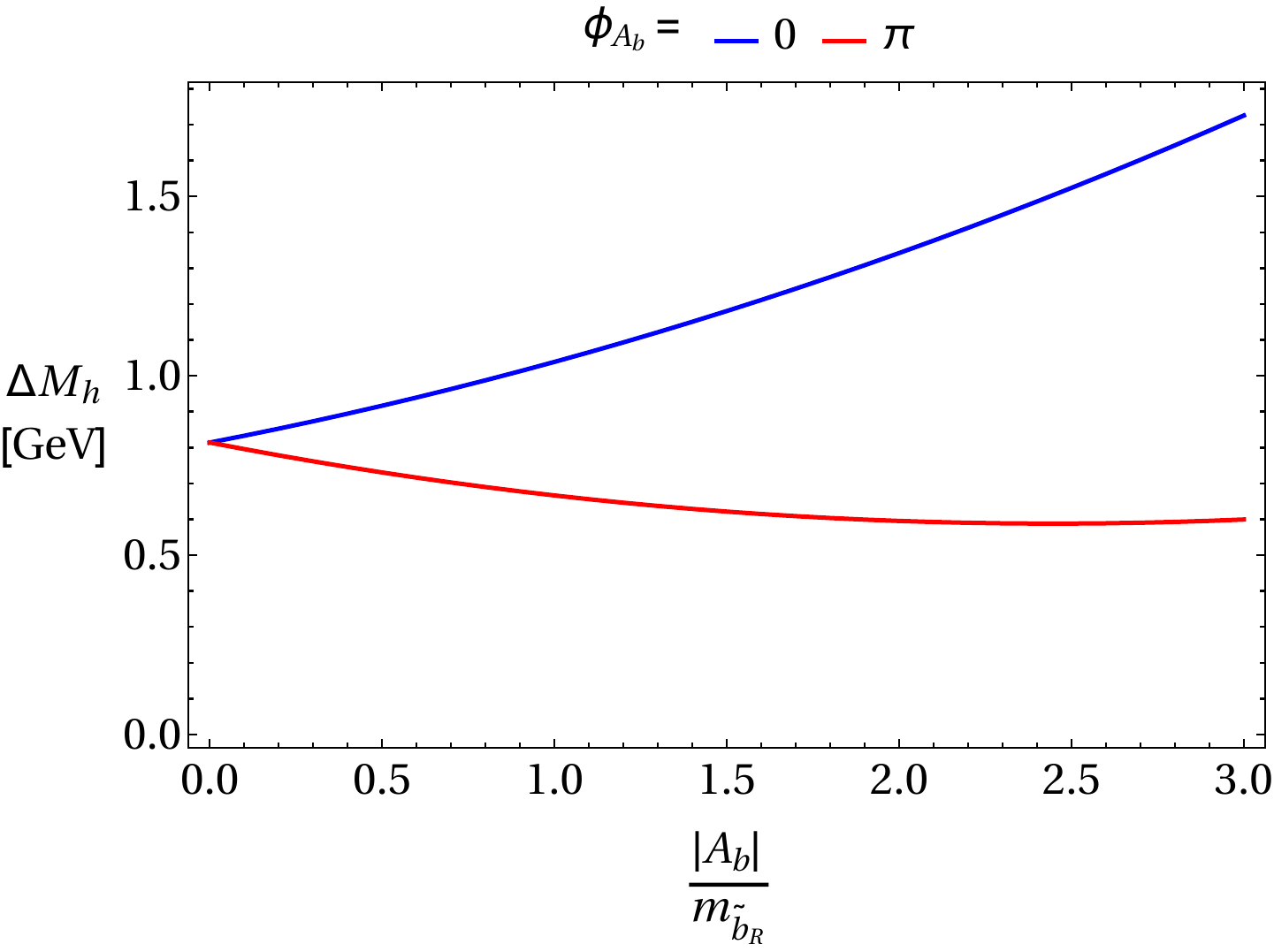}
  \caption{\label{fig:Ab}Dependence of the lightest Higgs-mass shift
    $\Delta M_h$ on $|A_b|$. The sign of $A_b$ is either positive
    (blue) or negative (red), and $\sgn{\mu} = -1$.}
  \end{minipage}
  \vspace{4ex}
\end{figure}

In our following analyses we choose~$m_{H^\pm}$ as an input parameter.
In this case
the~$\mathcal{O}{\left(\alpha_t\alpha_b+\alpha_b^2\right)}$ terms are
new contributions.  We investigate the dependence of the prediction
for $M_h$ on~$t_\beta, \mu$ and~$M_3$, whereby all parameters are
still kept real. The results are depicted in
Figs.~\ref{fig:tb}--\ref{fig:Ab}.

As can be seen in Fig.~\ref{fig:tb} large contributions above $1$\,GeV
are only visible at high values of $t_\beta$. In this scenario $M_3$
is positive, leading to a much bigger $\Delta M_h$ if $\mu$ is
negative, which can be understood from Eqs.~\eqref{eq:deltamb}
and~\eqref{eq:db}. For later analyses we fix $t_\beta = 50$.

In Fig.~\ref{fig:mu} we investigate the dependence of $\Delta M_h$ on
the size of $\mu$. We find very large gradients for the following two
cases: positive $M_3$ and negative $\mu\approx-1.8$\,TeV, and negative
$M_3$ and positive $\mu\approx2.6$\,TeV, which can again be understood
from Eqs.~\eqref{eq:deltamb}--\eqref{eq:db}, where for too large
values of~$\lvert\mu\rvert$ and opposite signs of~$\mu$ and~$M_3$ the
perturbative region of parameter space is left, as $\Delta
m_{b}\rightarrow -1$. A further increase of $\lvert\mu\rvert$ in the
regions of large gradients leads to a very strong enhancement of the
bottom Yukawa coupling and accordingly to very large negative mass
shifts, yielding eventually a tachyonic Higgs boson. For the following
analyses, we choose to fix $\mu=-1$\,TeV, \IE~below the problematic
scale and with $\sgn{\mu}=-1$. However, it should be noted that
scenarios with positive $\mu$ can lead to large shifts as well,
when~$M_3$ is negative, as in both cases the bottom Yukawa coupling is
enhanced. Moreover, scenarios with $\sgn{\mu}=1$ are in better
agreement with constraints from the anomalous magnetic moment of the
muon~\cite{Moroi:1995yh,Martin:2001st,Stockinger:2006zn}. Close to
$\lvert\mu\rvert = m_{\tilde{t}_{1,2}} - m_t \approx 1.8$\,TeV one can
see kinks which are induced by threshold effects from the
higgsino--top--stop system.

In Fig.~\ref{fig:M3} the impact of the gluino mass parameter is
depicted. This effect enters the Higgs self-energies at the
investigated order purely via the employed effective bottom mass. We
see a rising shift at growing~$\lvert M_3\rvert$ for opposite signs of
$\mu$ and $M_3$ (yielding the same enhancement in~$\Delta m_b$, see
Eqs.~\eqref{eq:db}). At \mbox{$\lvert M_3\rvert = m_{\tilde{b}_{1,2}}
  - m_b \approx 2$\,TeV} (nearly invisible) threshold effects from
the gluino--bottom--sbottom system appear. For our following analyses
we fix $\lvert M_3\rvert$ above that region at $2.5$\,TeV.

Finally, in Fig.~\ref{fig:Ab} the absolute value of~$A_b$ is varied,
and the resulting mass shift is plotted for positive sign (blue) and
negative sign (red) of~$A_b$. The difference between both curves,
\IE~the impact of the phase~$\phi_{A_b}$, is enhanced for larger
absolute values. However, as too large values of~$\lvert A_b\rvert$
lead to instable vacua according to the upper limit of
Eq.~\eqref{eq:ablimit}, we set it to $\lvert A_b\rvert = 2.5\,
m_{\tilde{b}_\text{R}}$ in the scenarios of the following section.

\subsection{Scenarios with complex parameters\label{sec:phases}}

\begin{figure}[t]
  \begin{center}
    \includegraphics[width=.9\linewidth]{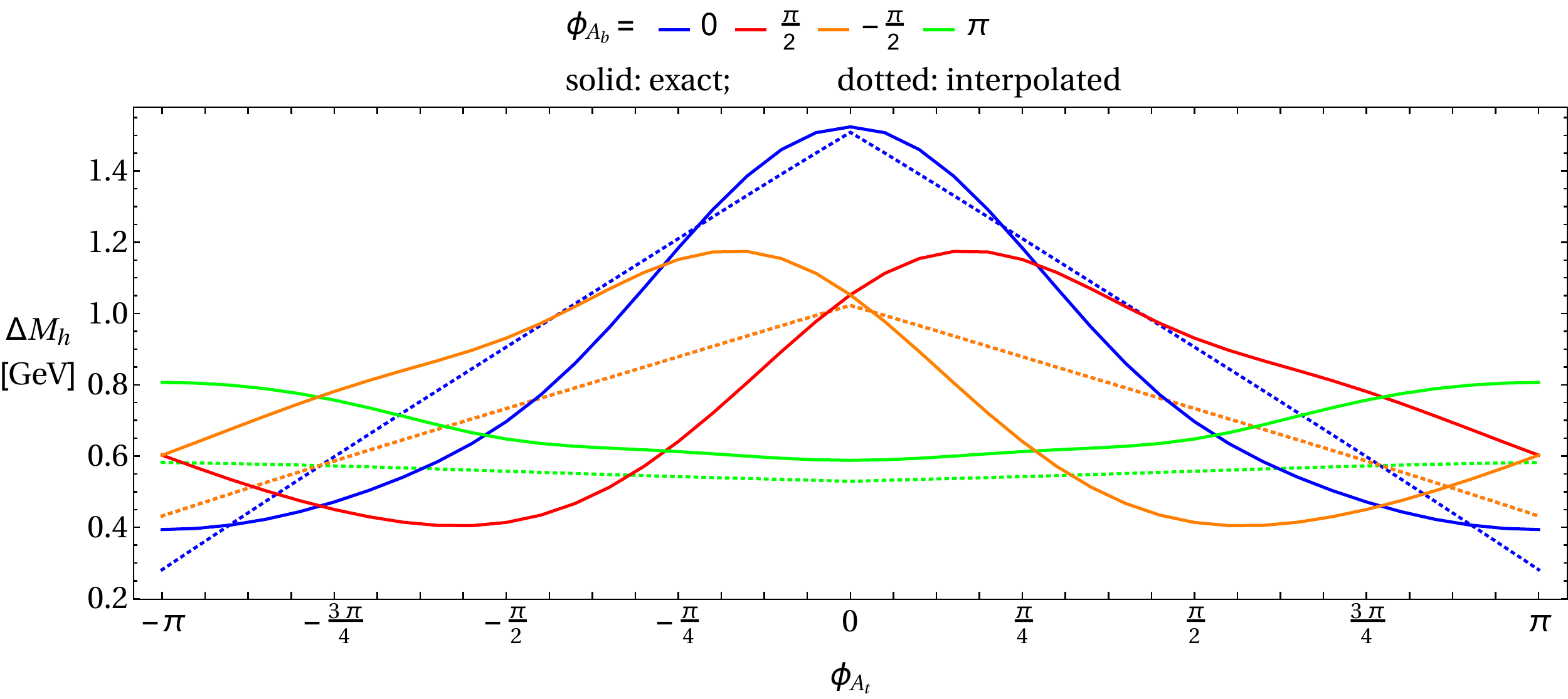}
    \caption{\label{fig:phiAt-phiAb}Dependence of the lightest
      Higgs-mass shift $\Delta M_h$ on $\phi_{A_t}$ and
      $\phi_{A_b}$, $\sgn{\mu} = -1$. solid: exact calculation,
      dotted: interpolation in \texttt{FeynHiggs}, the red-dotted and
      orange-dotted lines are identical.}
  \end{center}
\end{figure}

\begin{figure}[t]
  \begin{minipage}{.49\linewidth}
    \vspace{.05cm}
    \includegraphics[width=\linewidth]{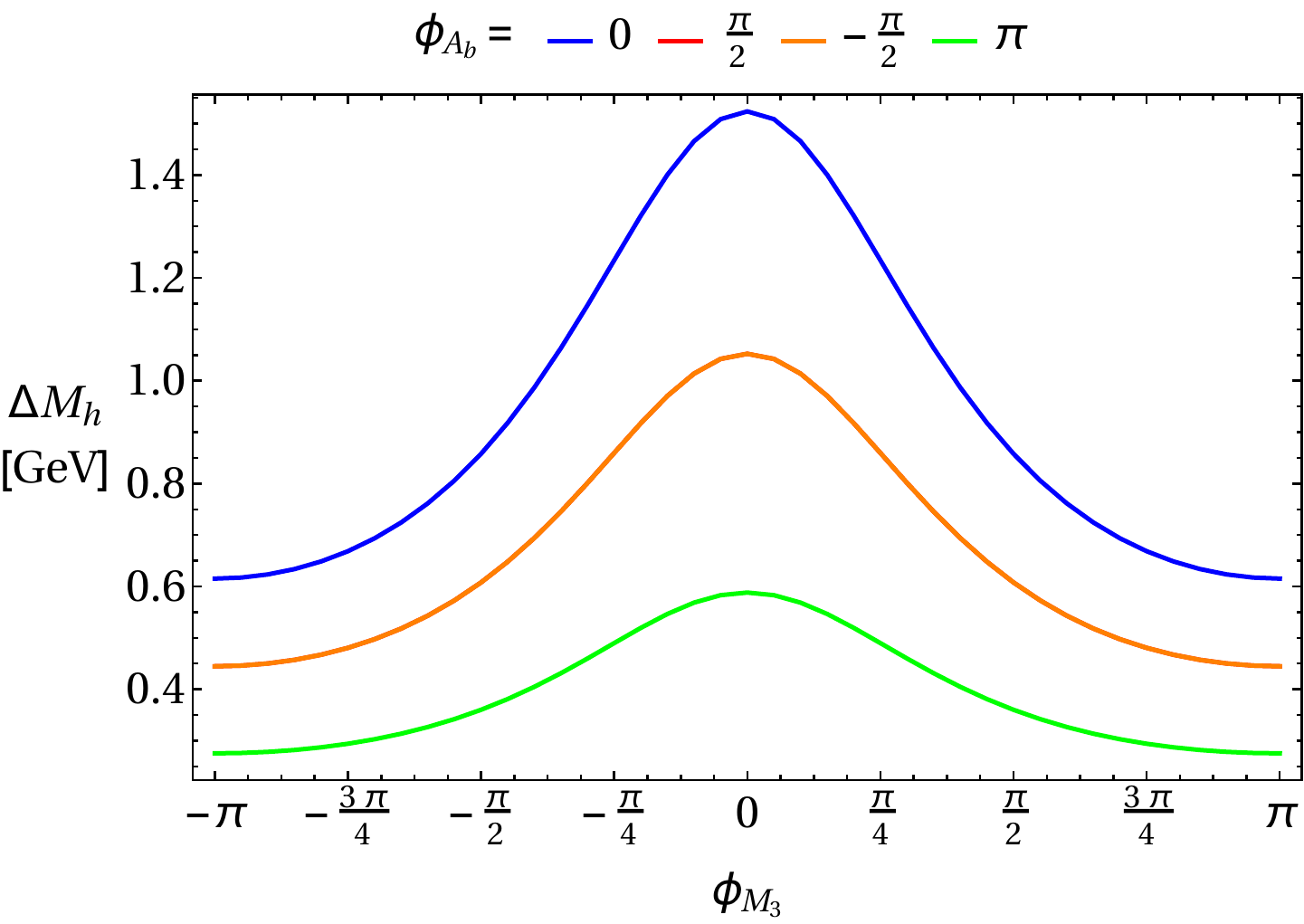}
    \caption{\label{fig:phiM3-phiAb}Dependence of the lightest Higgs-mass shift
      $\Delta M_h$ on $\phi_{M_3}$ and $\phi_{A_b}$, $\sgn{\mu} = -1$.}
  \end{minipage}
  \hfill
  \begin{minipage}{.49\linewidth}
    \includegraphics[width=\linewidth]{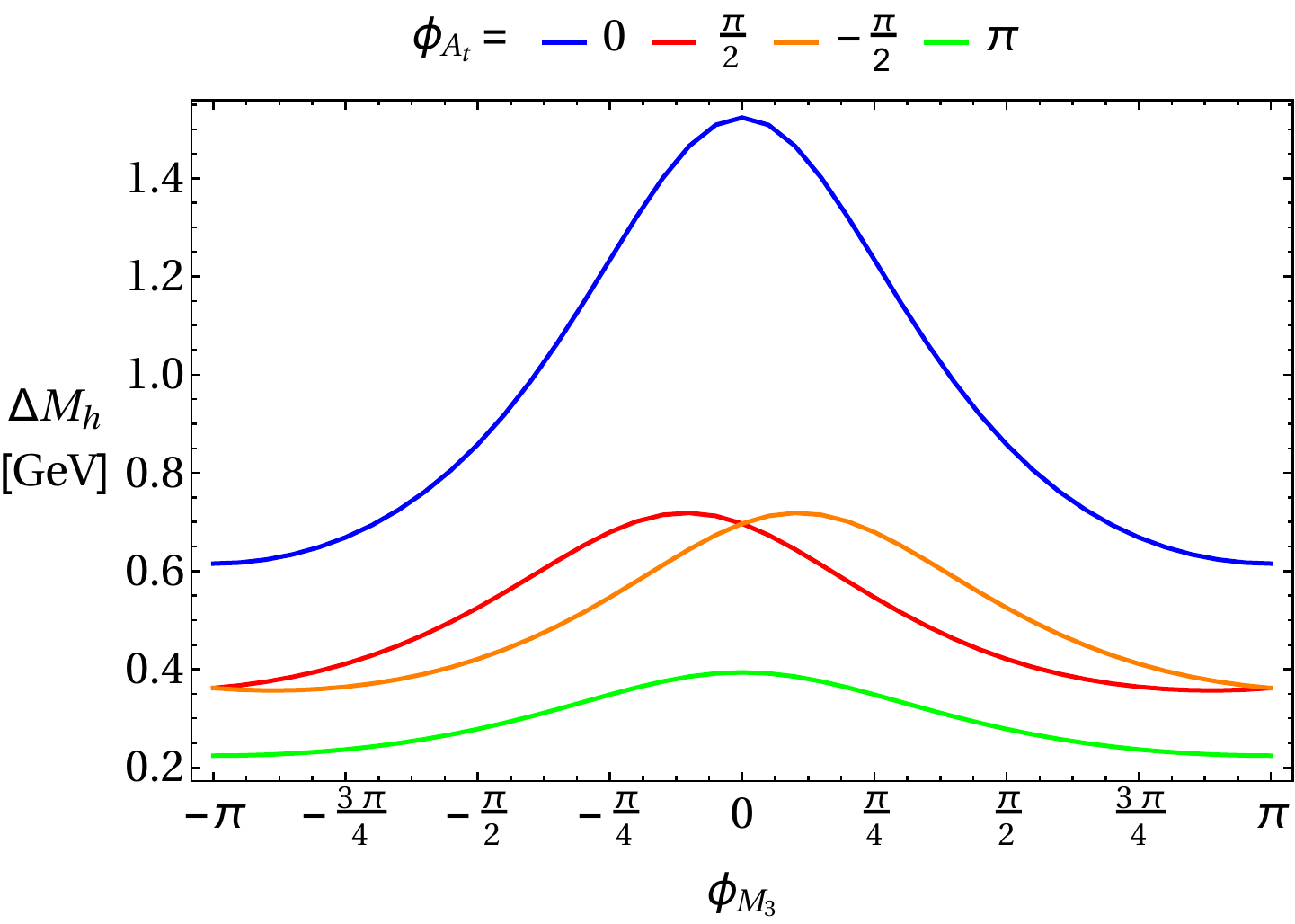}
    \caption{\label{fig:phiM3-phiAt}Dependence of the lightest Higgs-mass shift
      $\Delta M_h$ on $\phi_{M_3}$ and $\phi_{A_t}$, $\sgn{\mu} = -1$.}
  \end{minipage}
\end{figure}

Various phases enter the self-energies of the Higgs bosons at
$\mathcal{O}{\left(\alpha_t^2+\alpha_t\alpha_b+\alpha_b^2\right)}$. Their
impact on the Higgs sector is shown in
Figs.~\ref{fig:phiAt-phiAb}--\ref{fig:phiM3-phiAt}. Here we keep $\mu$
negative, \IE~$\sgn{\mu}=-1$, and $M_3$ positive, but we could also
have chosen the opposite signs of both parameters to see enhanced
effects for the phase dependent terms as has been shown in
Fig.~\ref{fig:mu}.

We start with the phases $\phi_{A_t}$ and $\phi_{A_b}$. The results
are depicted in Fig.~\ref{fig:phiAt-phiAb}, where mass shifts between
$0.3$\,GeV and~$1.4$\,GeV can be seen. For $\phi_{A_b} = 0$ the
variation with respect to~$\phi_{A_t}$ is maximal; the larger the
phase of $A_b$, the flatter the dependence on the phase
of~$A_t$. Similarly, variation of $\phi_{A_b}$ yields the largest
effects for $\phi_{A_t} = 0$. Also the signs of the phases matter,
\EG~the mass shifts are different for $\phi_{A_b}=\pm\tfrac{\pi}{2}$.

In addition to the exact calculation (solid lines), \texttt{FeynHiggs}
offers an implemented interpolation of the self-energy corrections
that have been known up to now for the case of real parameters but not
for the complex case. Since the
$\mathcal{O}{\left(\alpha_t\alpha_b+\alpha_b^2\right)}$ terms were
only available in \texttt{FeynHiggs} for the MSSM with real parameters
in the limits~$t_\beta\to\infty$ and~$m_b\to 0$, deviations from the
new mass shifts can be expected even for real parameters. Besides
these relatively small differences, the linear interpolation can
differ by $\approx 0.5$\,GeV from the full result in the investigated
scenario. Also the asymmetric behavior for the change of two phases at
the same time was not described correctly by the interpolation.

Figs.~\ref{fig:phiM3-phiAb} and~\ref{fig:phiM3-phiAt} show the
influence of varying the gluino phase $\phi_{M_3}$ and in addition
either $\phi_{A_b}$ or $\phi_{A_t}$. These terms are induced by the
correction factor~$\Delta m_b$ as the investigated class of two-loop
corrections does not contain the parameter $M_3$. Also here the
largest phase dependence is found when one phase is equal to zero. In
Fig.~\ref{fig:phiM3-phiAb} the mass shift is nearly symmetric
in~$\pm\phi_{M_3}$ and~$\pm\phi_{A_b}$, \IE~the red and yellow curves
are lying on top of each other. Nevertheless, there are small
asymmetries in the renormalized two-loop self-energies
$\hat{\Sigma}_{hA}$ and $\hat{\Sigma}_{HA}$. On the contrary the mass
shift~$\Delta M_h$ in Fig.~\ref{fig:phiM3-phiAt} shows a clear
asymmetry similar to Fig.~\ref{fig:phiAt-phiAb}.

In summary, phase dependent contributions of
$\mathcal{O}{\left(\alpha_t\alpha_b+\alpha_b^2\right)}$ lead to mass
shifts of the lightest Higgs boson of $\approx 1$\,GeV in the
investigated scenarios. The sign of $\mu$ has been chosen to be
negative in the considered scenarios, but similar effects can be found
at positive large $\mu$ (and opposite sign of $M_3$).

\section{Conclusions}

The two-loop corrections of
$\mathcal{O}{\left(\alpha_t^2+\alpha_t\alpha_b+\alpha_b^2\right)}$ to
the Higgs-boson masses in the MSSM with complex parameters have been
computed in the gauge-less limit at vanishing external momentum. The
terms of $\mathcal{O}{\left(\alpha_t\alpha_b+\alpha_b^2\right)}$ have
only been known in the special case of the MSSM with real parameters
before, and were incorporated in~\texttt{FeynHiggs} in the
limits~$t_\beta\to\infty$ and~$m_b\to 0$. The specific aspects related
to the renormalization of these new contributions have been discussed,
and their numerical impact on the Higgs spectrum has been
investigated.

For the lightest Higgs boson mass at $\approx125$\,GeV we have found
shifts above $1$\,GeV at $t_\beta>40$ for different scenarios:
moderate $\lvert\mu\rvert=1$\,TeV with negative sign and positive
$M_3$, $A_t$, $A_b$, or with positive sign and negative $M_3$, $A_t$,
$A_b$. The reason for that enhancement can be found in the large
correction factor~$\Delta m_b$ yielding an enhancement of the bottom
Yukawa coupling. The effect of varying the phases $\phi_{M_3}$,
$\phi_{A_t}$ and~$\phi_{A_b}$ can be as large as $1$\,GeV. If one
phase is set close to $\pi$, the dependence on the other phases is
typically weakened; the largest effects are found when only one phase
is varied with all others being zero. In \texttt{FeynHiggs} so far an
interpolation of the corrections
of~$\mathcal{O}{\left(\alpha_t\alpha_b+\alpha_b^2\right)}$ obtained
for the case of real parameters is used for the case of complex
parameters. We have found deviations with a size of~$\approx 0.5$\,GeV
from this approximation, especially when several phases are different
from zero at the same time.

Mass shifts for the heavier neutral Higgs bosons have not been
depicted. They are similar to the ones of the lightest Higgs boson,
however with opposite sign. Since we used a large value for~$t_\beta$
in our scenarios, we need to choose a rather large input
mass~$m_{H^\pm}=1.5$\,TeV to be consistent with existing experimental
bounds. Therefore, the relative size of the mass shifts is
small. Moreover, both heavy Higgs bosons receive similar corrections
with a maximal difference of~$\approx 0.1$\,GeV in the investigated
scenarios. Nevertheless, small mass shifts can be important to
correctly describe the resonance-type behavior of nearly
mass-degenerate mixed states like the two heavy Higgs bosons in the
MSSM with complex parameters.

The new results will be implemented in the public
code~\texttt{FeynHiggs}.

\section*{Acknowledgments}

We thank H.~Bahl, T.~Hahn, S.~Heinemeyer, W.~Hollik, W.~G.~Hollik and
D.~Stöckinger for helpful discussions. Special thanks go to P.~Slavich
for help in comparisons with his code. This work has been supported by
the Collaborative Research Center SFB676 of the DFG, ``Particles,
Strings and the early Universe'', and by the European Commission
through the ``HiggsTools'' Initial Training Network
PITN-GA-2012-316704.

\begingroup
\bibliographystyle{h-physrev}     % Zitierstil: alpha = [Nam88]
\setstretch{.5}
\bibliography{literature}
\endgroup

\end{document}